\documentstyle[12pt,aas2pp4]{article}
\def\etal{{\it et al.\/}}

\def\lae{\mathrel{<\kern-1.0em\lower0.9ex\hbox{$\sim$}}}
\def\gae{\mathrel{>\kern-1.0em\lower0.9ex\hbox{$\sim$}}}
\def\flux{{erg s$^{-1}$cm$^{-2}$\AA$^{-1}\/$}}
\slugcomment{Submitted to the  Astrophysical Journal}
\begin{document}
 
\title{Ultraviolet Imaging Observations of the cD Galaxy in Abell 1795:Further
Evidence for Massive Starformation in a Cooling Flow}
 
\author{Eric P. Smith}
\affil{Laboratory for Astronomy and Solar Physics, Code 681 - NASA/Goddard
Space Flight Center,Greenbelt, MD 20771}
\authoremail{esmith@hubble.gsfc.nasa.gov}

\author{Ralph C. Bohlin}
\affil{Space Telescope Science Institute, Homewood Campus, Baltimore, MD 21218}

\author{G. D. Bothun}
\affil{Department of Physics \& Astronomy, University of Oregon, Eugene, OR
97403}
 
\author{Robert W. O'Connell}
\affil{Astronomy Department, University of Virginia, Charlottesville, VA
22903}
 
\author{Morton S. Roberts}
\affil{National Radio Astronomy Observatory, Edgemont Road, 
Charlottesville, VA 22903}
 
\author{Susan G. Neff, Andrew M. Smith \& Theodore P. Stecher}
\affil{Laboratory for Astronomy and Solar Physics, Code 680 - NASA/Goddard
Space Flight Center,Greenbelt, MD 20771}

\clearpage

\begin{abstract}
We present images from the Ultraviolet Imaging Telescope of the Abell 1795 
cluster of galaxies.  We compare the cD galaxy morphology and photometry of these 
data with those from existing archival and published data.  The addition of 
a far--UV color helps us to construct and test star formation model 
scenarios for the sources of UV emission.  Models of star formation with 
rates in the range $\sim5-20M_{\sun}$yr$^{-1}$ indicate that the best fitting 
models are those with continuous star formation or a recent ($\sim4$ Myr old) 
burst superimposed on an old population.  The presence of dust in the 
galaxy, dramatically revealed by HST images complicates the interpretation 
of UV data.  However, we find that the broad--band UV/optical colors of 
this cD galaxy can be reasonably matched by models using a Galactic form 
for the extinction law with $E_{B-V}=0.14$.  We also briefly discuss other 
objects in the large UIT field of view.
\end{abstract}


\section{Introduction}            \label{sec:introduction}

The final fate of the cooling gas seen by X--ray telescopes in clusters of 
galaxies has long been sought and any system which can provide evidence for 
star formation arising in a system of cooling gas is of great interest. The 
cD galaxy in Abell 1795 ($z=0.0634$) has been an object of much study since 
it was identified with the radio source 4C 26.42 (\cite{mer72}) and later 
found to reside in a cluster cooling flow (\cite{mck80}).  Subsequent 
optical studies (\cite{sar73,hec81}) found it to possess extended nuclear 
emission--line gas ($d\sim20$kpc, $H_o=75, q_o=0.1$).  Spectroscopic 
studies in both the ultraviolet and optical reveal an unusual blue 
continuum (\cite{emh92,all95}) whose structure has been recently traced in 
broad band optical imaging (\cite{mcn92}).  These properties have made it 
one of the best candidate galaxies for exhibiting star formation arising 
from a cluster cooling flow though other mechanisms can be invoked to 
explain the radio emission and emission--line gas.  

Hubble Space Telescope observations, improved CCD sensitivity at blue 
wavelengths and the ultraviolet telescopes on board the Astro Space Shuttle 
payload have made this galaxy an object of renewed interest 
(\cite{mcn93,mcn96,pin96}).  These recent observations have revealed 
regions with colors bluer than normal ($(U\!-\!I)\sim2.1$;\cite{mcn93}) for 
an expected old stellar population, near the galaxy nucleus that may be the 
sites of recent star formation.  These blue regions are confined to the 
very central ($r\la 20$\arcsec\/) region of the galaxy.  The cD appearance 
at long wavelengths ($\lambda > 7000$\AA\/) where the light is dominated by 
the oldest stars is smooth and does not suggest that a recent globally 
disturbing event (e.g.  a major galaxy interaction) has occurred to trigger 
a burst of central star formation.  Suggested possible triggers for the 
star formation have included radio jets, collision with a smaller galaxy 
(\cite{mcn96}), and the interaction of the gas cooling out of the X--ray 
halo with the galaxy interstellar medium (ISM).  Each of these 
interpretations predict some recent star formation in the galaxy nucleus, 
but with potentially differing morphologies, resultant population mixes and 
emission--line properties.

If the galaxy has undergone recent star formation then it is natural to 
turn to the ultraviolet (UV) part of the spectrum for observational 
confirmation.  Young stars will most easily be distinguished from the 
preexisting stellar population by observations shortward of the 4000\AA\/ 
break.  In particular, observations by the Ultraviolet Imaging Telescope 
(UIT) in its far--UV filter may be used as a valuable diagnostic for stars 
which are extremely hot ($T_{eff} \gae 10000$K).  At the distance of cD 
($z=0.0634$) as few as 1200 O stars would be detectable with UIT (provided 
there was no obscuration by dust).  At the wavelengths observed by the 
UIT all of the underlying old population (primarily G, K and later type stars) 
will be invisible, with the important possible exception of extreme 
horizontal branch (EHB) stars (\cite{fer91,dor95}).

In section~\ref{sec:data} we discuss the UIT data and photometry as well as
archival data used in the analyses. We follow in section~\ref{sec:summary} 
with a discussion of the implications these new UV data have for star 
formation properties in Abell 1795.

\section{Observational Data}      \label{sec:data}

We present our image obtained with the UIT during the Astro--2 Space 
Shuttle mission of March 1995 in figure~\ref{fig:cluster}.  The UIT took 
several exposures using a far--UV filter with $\lambda_c$=1520\AA\/; 
$\Delta\lambda=356$\AA\/ (which we shall hereafter denote by the first two 
digits of its central wavelength).  For this study we used only the deepest 
(1310.5 sec) exposure.  For a complete description of the instrument and preliminary data 
reduction steps involved see \cite{ste92}.  The UIT has a large field of 
view (40\arcmin diameter) so the entire cluster core was imaged, but little 
beyond the cD itself was detected.  The nominal stellar point--spread 
function (PSF) for the UIT has a FWHM of $\sim3$\arcsec, but the Abell 1795 
images have poorer resolution with FWHM of the stellar PSF of $\sim6.8$ 
\arcsec.  Several factors can contribute to decreased resolution with the 
most common being crew motions onboard the shuttle during the observation.  
In section {\ref{sec:others}} we briefly discuss objects other than the cD in the 
UIT image.

We have also assembled a collection of archival data to complement our UIT 
observations of this cluster.  These additional data, when combined with 
that from UIT create a unique, long wavelength lever arm with which to 
examine the star formation properties.  We extracted two pipeline 
calibrated image data sets, taken with the F555W (HST ``V'') and F702W filters, from 
the Hubble Space Telescope (HST) archives (u2630401-6).  
After correcting the image headers for updated WFPC2 astrometry (J. Hu, 
private communication) we removed the cosmic rays and applied a charge 
transfer correction as outlined in \cite{hol95} and combined the two 
longest exposure images (800 sec each).  Since the portion of the cD galaxy 
seen by the UIT is the central $\sim20$\arcsec\/ we used only that portion 
of the HST data (PC chip).  No sky subtraction was performed on the HST 
images because the short exposure times (800 sec) combined with small pixel 
size imply that there is little if any contribution to the F555W and F702W 
signal from a background.

B. McNamara kindly provided us with a copy of his calibrated $U$--band image
obtained with the Kitt Peak National Observatory's 2.1m telescope
(\cite{mcn92}). For comparing the global stellar population with that of 
the electrons responsible for the radio emission we acquired a copy of 
previously published VLA data from J.P. Ge (\cite{jpg93}).

\subsection{Morphology}           \label{sec:morphology}

One of the primary goals of the UIT was to investigate the changes in 
galaxy morphology with wavelength for a variety of galaxy types.  As part 
of this effort we have constructed a series of figures comparing the UIT 
image of the cD galaxy with those from other passbands 
(figures~\ref{fig:uitha} through \ref{fig:uitvla}).  These figures show the 
UIT image resembles the H$\alpha$ emission more nearly that the broad band 
optical emission and that the UV morphology has no obvious link to the 
radio morphology. The similarity of the UV
and H$\alpha$ morphologies prompted us to estimate the amount of Ly$\alpha$
contamination in our UIT filter band.  To determine 
the strength of any Ly$\alpha$ emission from the center of the galaxy we 
extracted an archival IUE spectrum of the cD galaxy (\cite{hcw85}) and 
convolved it with the UIT filter response curve.  Though the system does 
exhibit moderately strong Ly$\alpha$ emission (see 
\S\ref{sec:photometry}) this contributes little to the UIT measured 
flux.  Moreover, Astro-2 Hopkins Ultraviolet Telescope 
spectra show those emission--lines that are present in the UIT filter are 
weak and do not contribute substantially to the overall UV flux (Dixon, 
private communication).  Thus, despite the superficial resemblance of the 
UIT cD image to the H$\alpha$ morphology we conclude that the light is from 
a continuum source and not extended emission--line gas. 

Direct comparison of the HST optical and UIT UV morphologies is complicated 
to interpret due to the great difference in resolution.  However, the 
presence of the dust lane dramatically revealed in the HST data 
(Fig. \ref{fig:hstuit}) is reflected in the pinching of the UV contours in the 
UIT image. In general, the UV morphology is much less regular than the HST 
F555W and F702W morphologies.  The blue lobes seen by \cite{mcn93} may be 
associated with the North and South peaks of the UV contours separated by 
$\sim10$\arcsec\/, but we do not find these regions to be bluer than the 
galaxy as a whole (in 1520\AA\/-$V\equiv$15-$V$\/). The 3.6cm and 1520\AA\/ 
morphologies of the cD galaxy are coaligned roughly North--South 
(figure~\ref{fig:uitvla}).  \cite{mcn93} found the blue lobes at the ends of the 
radio emission leading them to posit that the jets may be inducing star 
formation via interaction with the galaxy ISM.  The UV data suggest very 
weak concentrations in flux near the outer edges of the radio emission.

To further investigate the morphology of the UV light we compare it surface 
brightness profile (Fig.  \ref{fig:profiles}) with those at longer 
wavelengths We generated surface brightness profiles in the UV, F555W and 
F702W bands using the STSDAS {\tt ellipse} task (Fig.  \ref{fig:rprofile}), 
and fit a function to the profile in \cite{mcn93} for the $U$\/ profile.  
There is no evidence for a point source component to the UV surface 
brightness profile.  We fit both elliptical ($r^{\frac{1}{4}}$--law) and 
disk galaxy (exponential) light profiles to the UIT data and find that the 
disk galaxy light profile is a better match.  Table~\ref{tab:galfits} gives 
the results for our model galaxy profile fits.  We also performed fits to 
the F702W profile.  The best fit was achieved with a combined point source, 
exponential disk and $r^{\frac{1}{4}}$-law bulge model.  However, the disk 
central surface brightness for this fit is more than 2 magnitudes brighter 
than the corresponding characteristic bulge surface brightness ($I_e$\/) 
indicating that the central region ($r < 8\arcsec$) of the cD has a 
predominantly disk-like light distribution.  Thus both the UV and visual 
light from the central portion of the galaxy are better characterized by 
exponential light distributions than by elliptical galaxy light 
distributions, but with the UV light being more extended 
($r\sim20\arcsec$).  Interestingly, \cite{hcw85} found spectroscopic 
evidence for for emission--line gas rotation in the central part of the cD 
while \cite{ant93} found no evidence for a rotating disk, but neither 
measurements were sensitive to the small spatial scales accessible to HST 
nor to the young stellar population easily isolated by the UIT.

\placetable{tab:galfits}

The \vr\/ color index does not vary significantly over the central portion 
of the cD.  However, the (15-$V$\/) index does change with the central 
part being slightly redder than the outer parts.  This is most likely do 
to the presence of the dust so easily visible in the HST image.  
\cite{oco92} found that several ellipticals observed with UIT generally 
become bluer with decreasing radius, but none of those galaxies were know 
to possess significant amounts of centrally concentrated dust.

\subsection{cD Galaxy Photometry}           \label{sec:photometry}

We performed our photometry using software specifically tailored to UIT 
data, but modeled on the DAOPHOT aperture photometry routine.  Curve of 
growth analysis shows that the cD galaxy light (centered at 
$\alpha=$13:48:52.42, $\delta=+$26:35:34.7) is contained within a circle of 
radius $\sim25$\arcsec.  The aperture magnitude and corresponding flux for 
the galaxy are $m_{15}=15.54\pm0.09$ ($2.21\times10^{-15}$\flux).  
Henceforth, all photometric comparisons will be confined to this region or 
specific apertures within this radius.  Using archival IUE spectra we are 
able to estimate the contribution to the UIT flux arising from the 
Ly$\alpha$ emission in the galaxy (see also \cite{emh92}).  We have 
corrected the UIT magnitudes by $\Delta\!m_{15}=0.08$ to account for this 
emission--line contribution, yielding $m({\rm 
continuum})_{15}\equiv\!m_{c,15}=15.62\pm0.09$.  There was no detectable UV 
flux associated with the blue filamentary region described by \cite{mcn93}.  
To check for the presence of the blue lobes we placed small 
($5\times5$\arcsec) apertures on the locations of lobes (from \cite{mcn93} 
figure 6) and on the galaxy center.  We find no significant difference 
between the colors in these various regions.  However, the low spatial 
resolution of the UIT image and typical uncertainty in its photometry 
($\sim0.1$mag) folded in with stochastic variations of the spatial 
distribution of O--stars imply that detecting color difference with these
images would be difficult.

We photometered the HST and U band data within the same region and list the 
resulting magnitudes and colors in table~\ref{tab:photometry}.  Steps 
outlined in \cite{hol95} were taken to calibrate the HST photometry.  For 
both the U and HST images we digitally removed a small elliptical galaxy at 
$r\sim9.5$\arcsec\/ replacing the pixels with those from a region at the 
same radius, but 180 degrees different in position angle.  The HST 
magnitudes were corrected for emission--line contributions using the data 
from \cite{ant93}.  These corrections were small with $\Delta\!V\/=0.02$ 
and $\Delta R\/=0.06$ respectively.  The only emission line in the 
$U$--band filter is the [\ion{O}{2}]$\lambda3727$ for which we correct by 
$\Delta U\/=0.04$.  The optical magnitudes have been $K$--corrected using 
the models from \cite{cww80}.  We estimated the $K$--correction for the 
1520\AA\/ ($K_{15}=0.26$) by convolving a numerically redshifted O star 
spectrum with the UIT filter response and comparing this with its $z=0$ 
counterpart.  No corrections for Galactic extinction have been applied 
because $E_{B-V}\la0.01$.

\placetable{tab:photometry}

We also photometered a polygonal region ($\sim8\times4$\arcsec) centered on 
the dust lane visible within the HST image to compare the colors inside and 
outside this region.  The change in color compared with an identical region 
centered on the optical nucleus is $\Delta({\rm 15}-V)=0.8$.  

\subsection{Other Objects}           \label{sec:others}

There are 10 previously cataloged objects, excluding the cD galaxy, visible 
in the UIT image.  Five of these are cataloged only in the HST Guide Star 
Catalog (GSC) while the other five are SAO82997 ($m_{15}=14.00\pm0.07$, 
(15-$V$\/)=4.89, F8V), SAO83004 ($m_{15}=14.30\pm0.10$, (15-$V$\/)=4.97, 
F0V), Abell cluster galaxies \#40 and \#151,(\cite{huc90}) and 1345.6+2639 
($m_{15}=15.85\pm0.10$, (15-$V$\/)=0.15; a Seyfert galaxy).  The two stars 
and the Seyfert galaxy are all point sources.  There is no previously 
published photometry for the two cluster galaxies, but we have estimated 
their colors based upon their digitized Palomar Sky Survey (DPOSS) images.  
We find: Abell \#40, $m_{15}=16.4\pm0.2$, (15-$V$\/)=3.0, Abell \#151, 
$m_{15}=16.6\pm0.2$, (15-$V$\/)=2.3.  These magnitudes have not been corrected 
for $K$--dimming or for extinction (external or intrinsic).  Both 
objects are quite blue for normal cluster galaxies and may be experiencing 
some star formation.  Assuming their UV light is due to young stars we 
estimate that Abell \#40 and Abell \#151 contain $\sim8500$ O stars.  This 
would imply star formation rates of approximately 1 M$_{\sun}$yr$^{-1}$ We 
note however that their (15-$V$\/) colors are similar to those seen in 
galaxies with ``UV upturns'' arising from hot EHB stars in the old 
population (\cite{dor95}).  Therefore, without other evidence for recent 
star formation (H$\alpha$ emission for example) we cannot distinguish 
between young stars and EHB stars as the primary cause for blue colors.

In addition to these sources there are four uncataloged sources (all are 
5$\sigma$ detections) with mean magnitude $<m_{15}>=16.7$.  Comparison of their
positions with the DPOSS suggest all of these are stars or groups of stars too
faint in the visible to be included in the  HST GSC.  Comparison with deep
optical images has proven to be fruitful in identifying faint UIT sources
(\cite{one96}) which are easily overlooked on the POSS.  We expect the same 
may be true for this image as there are several faint ($2-3\sigma$\/) 
sources without optical counterparts in the DPOSS.

\section{Discussion and Summary}  \label{sec:summary}

To estimate the star formation history or current star formation rate using 
the UV measurements from UIT we must first determine the nature of the UV 
light.  If the light arises primarily from emission--line gas then it will 
tell us little about massive star formation.  Early IUE spectroscopy by 
\cite{nor84} suggested that the UV light was due to high temperature gas, 
but later IUE observations did measure a continuum (\cite{emh92,cra93}).  
Moreover, recent optical spectroscopy has found strong evidence of a hot 
continuum (\cite{all95}) at optical wavelengths.  Spectroscopy at all 
wavelengths reveals line ratios from the emission--line gas that are 
inconsistent with ionization via a nuclear, nonthermal power--law source.  
The LINER--like spectrum of Abell 1795 is better fit by shock models 
(\cite{wvb84,emh92,ant93}) or photoionization by a hot population 
(\cite{all95}).  Thus, it seems likely that the bulk of the UV emission in 
the UIT bandpass does come from a continuum of hot stars which may in turn 
be responsible for ionizing some of the emission--line gas.

\subsection{Star Formation}          \label{sec:models}

Since at least some portion of the observed H$\alpha$ and Ly$\alpha$ flux 
arises from gas photoionized by a UV continuum there must be a significant 
population of young hot stars, implying either a recent episode of star formation 
($t\sim$few Myr) or continuous star formation.  The UIT magnitude listed in 
 Table \ref{tab:photometry} implies a luminosity at 1500\AA\/, 
$L_{1500}=1.9\times10^{42}$ erg sec$^{-1}$.  We can estimate the number of 
O5 stars required to produce the measured UIT flux and then compare that 
with the expected star formation rate needed to fuel the H$\alpha$ 
emission.  Using Kurucz model atmospheres (\cite{kur93}), assuming and 
effective temperature and luminosity of 38,000K, $\log(\frac{L_{bol}}{L_{\sun}})=5.7$ for 
a mid O star, and cosmology with $H_o=75$, $q_o=0.1$ we require 
$1.8\times10^4$ stars to match the UIT, $K$--corrected flux.  This 
represents the {\em minimum\/} number of ionizing stars because no extinction 
correction has been applied and mid O stars produce more ionizing flux than 
early B stars.  Without ultraviolet spectroscopic information we cannot 
further restrict the relative contributions of various stellar types with 
temperatures greater than 15000K to the UIT flux.  Adopting the extinction 
($E_{\bv}=0.14$) from \cite{emh92} and assuming a Galactic extinction law 
(Savage and Mathis 1979) and foreground screen dust model causes the 
estimated number of O stars to rise to $5.3\times10^4$.  Recent optical 
spectroscopy by \cite{all95} predicts $2.4\times10^4$ O stars (adjusting to 
our cosmology) using similar extinction, suggesting that the screen model 
may be a reasonable approximation for the dust extinction in the galaxy 
center.  The implied star formation rate (SFR) is in the range 
$8-23M_{\sun}$yr$^{-1}$, depending on the IMF model, with the lower bounds 
corresponding to a \cite{ken83} IMF and the upper bound corresponding to a 
\cite{mil79} IMF. Since some of the H$\alpha$ emission comes from shocked 
gas a SFR derived assuming all the emission come from photoionized gas is 
an upper limit.  Therefore, it is likely that the actual SFR is closer to 
the lower limit in the case of Abell 1795.  This SFR is consistent with the 
predicted star formation rates one calculates from the H$\alpha$ luminosity 
($5M_{\sun}$yr$^{-1}$, \cite{wvb84,ken83}). These star formation rates are
substantially lower than those estimated from recent X--ray analyses 
($\sim300M_{\sun}$yr$^{-1}$; \cite{edg92,fab94}).

To further test the various scenarios for possible star formation histories 
we use models based upon (a) a single, instantaneous burst of star 
formation (b) a single, exponentially decaying burst of star formation (c) 
continuous, constant star formation (d) a single, instantaneous old burst 
with smaller, younger bursts superimposed.  The first three models are 
described in more detail in \cite{cor94}.  The fourth model is simply a 
scaled superposition of the models in (a).  We have plotted color--color 
diagrams using the four colors (15-$V\!$), (15-$U\!$), ($U\!-V\!$), (\vr) 
for each set of the above models to compare with the global colors of the 
cD galaxy (figure~\ref{fig:c-cd}).  The colors have been corrected for 
$K$--dimming, but not extinction.  Extinction vectors based upon various 
models for the distribution of the dust are added to the plot.  Several 
models for the star formation history can be ruled out.  Models with 
exponentially decaying star formation and those in which the central 
population was created in a single burst of age greater than $\sim$few 
hundred Myr all have broad-band colors too red to match the cD. Models with 
some levels of continuous star formation and/or those with recent bursts 
however can reproduce the colors.  The most plausible burst model is one in 
which an old galaxy ($z_{form}=5$) experiences a Salpeter IMF burst of age 
4Myr and total mass $25\%$ of the old population, within the photometered 
aperture.  We estimated the mass interior to $r=\sim25$\arcsec\/ by 
comparing the fraction of light within this radius for an 
$r^{\onequarter}$--law galaxy with its total and assumed a constant $M/L$ 
ratio of 20.  Since this likely underestimates the central $M/L$ ratio the 
mass required of our burst to match the cD colors is an upper limit.  The 
most plausible continuous, constant star formation model which matched the 
far UV - visible colors is one inn which the galaxy has been forming stars 
at a rate $\sim5-10M_{\sun}$yr$^{-1}$ over the past 5 Gyr.

\subsection{Dust Extinction}      \label{sec:dust}

The dust lane visible in figure~\ref{fig:hstuit} complicates the 
interpretation of broad--band observations, particularly those involving 
ultraviolet light.  By comparing the measured $IUE$ Ly$\alpha$ flux to 
H$\alpha$ flux and assuming suitable values for the intrinsic line flux 
ratio, \cite{emh92} estimated the intrinsic $E_{\bv}$=0.14, but with a wide 
possible range ($0.2 < E_{\bv} < 0.22$).  Recent optical spectroscopy by 
\cite{all95} suggests $E_{\bv}$=0.22.  The straight line vectors in 
figure~\ref{fig:c-cd} show the effects of dust assuming both the Galactic 
\cite{sav79} 
($A_{1520}=8.32E_{\bv}$) and SMC extinction laws.  The curved lines (for a 
dusty galaxy, similar to a spiral and an elliptical galaxy model) are the 
extinction curves based upon the dust/stellar distribution models of 
\cite{wit92}.  Their models of mixed dust, gas and stars generally produce 
smaller extinction in the UV than a simple foreground screen models.  The 
location of the galaxy in the color--color diagrams is consistent with 
systems having continuous star formation over the last 5-10Gyr when 
combined with the $E_{\bv}$=0.14, Galactic-law screen or dusty galaxy 
extinction models, or with a system characterized by an old population with 
a 2-5Myr burst and the same extinction laws.  Extinction laws with 
$E_{\bv}$=0.22 combined with measured cD colors generally produce values 
too blue to be consistent with any of our galaxy models.

\subsection{Conclusions}

Our analyses reveal several things.  The light at 1520\AA\/ is best 
characterized by an exponential or disk-like light profile with a scale 
length of 4kpc.  Similarly, the F702W light is best matched by a 
predominantly disk-like profile but with scale length of 2 kpc.  The 
UV--optical colors match those of models with continuous star formation 
for $t\sim 5-10$ Gyr or those of an old galaxy with a massive, young 
($t\sim4$ Myr) burst.  Both of these properties are consistent with the UV 
light coming from an accreted population formed in a cluster cooling flow 
or from an episode of star formation induced by an interaction with a much 
smaller galaxy.  The cooling flow scenario has the following points as 
supporting evidence (a) line ratios indicative of shock heated gas and (b) 
the inner surface brightness profile does not show significant disturbances 
in the UV or optical bands.  However, detailed image analyses of deep $U$\/ 
imaging by \cite{bri96} appears to have revealed possible debris associated 
with an interaction of the cD with two small galaxies.  These features are 
too faint to have been detected with the UIT. High resolution images in the 
UV would help further constrain the nature of the UV bright population.

\acknowledgements  We wish to thank Brian McNamara for providing us with 
a $U\/$-band image of Abell 1795.  EPS also thanks 
Jason Pinkney for helpful comments about photometry of HST images.  This
research has made use of the NASA/IPAC Extragalactic Database (NED) which is
operated by the Jet Propulsion Laboratory, Caltech, under contract with the
National Aeronautics and Space Administration.
 
\clearpage

\begin{deluxetable}{lrr}
\tablecaption{Surface Brightness Profile Models} \label{tab:galfits}
\tablewidth{0pt}
\tablehead{
\colhead{Model} & \colhead{Scale Length [\arcsec]} & \colhead{$\chi^2$}}
\startdata
UV: Disk                                       & $ 4.5\pm 0.3$  &5.4\nl
UV: $r^{\frac{1}{4}}$-law                      & $41.1\pm 16.6$ &18.3\nl
$R$\/: Disk \tablenotemark{a}                  & $ 3.7\pm0.04$  &4.9\nl
$R\/ : r^{\frac{1}{4}}$-law \tablenotemark{a}  & $142\pm32$     &42.7\nl
$R$\/: (composite) Disk+$r^{\frac{1}{4}}$-law+point source &                &0.54\nl
$R$\/ (composite) Disk component                           & $2.2\pm0.3$    &\nl
$R$\/ (composite)$r^{\frac{1}{4}}$-law component          & $0.3\pm0.1$          &\nl
\enddata
\tablenotetext{a}{Fits performed over $ 0.3\arcsec < r < 
8\arcsec$ region of profile to exclude central peak in HST profile.}
\end{deluxetable}

\clearpage

\begin{deluxetable}{lr}
\tablecaption{Photometric Properties\tablenotemark{a}} \label{tab:photometry}
\tablewidth{0pt}
\tablehead{
\colhead{Parameter} & \colhead{Value} }
\startdata
$m_{c,15}$ & $15.36\pm0.07$\nl
(15-$U$\/) & $-0.18 $ \nl
(15-$V$\/) & $0.53 $  \nl
(\vr)      & $0.61 $ \nl
\enddata
\tablenotetext{a}{All properties measured within $r=12$\arcsec\/ aperture.}
\tablenotetext{b}{Values include $K$\/-corrections (see~\ref{sec:photometry}).}
\end{deluxetable}

\clearpage

\figcaption[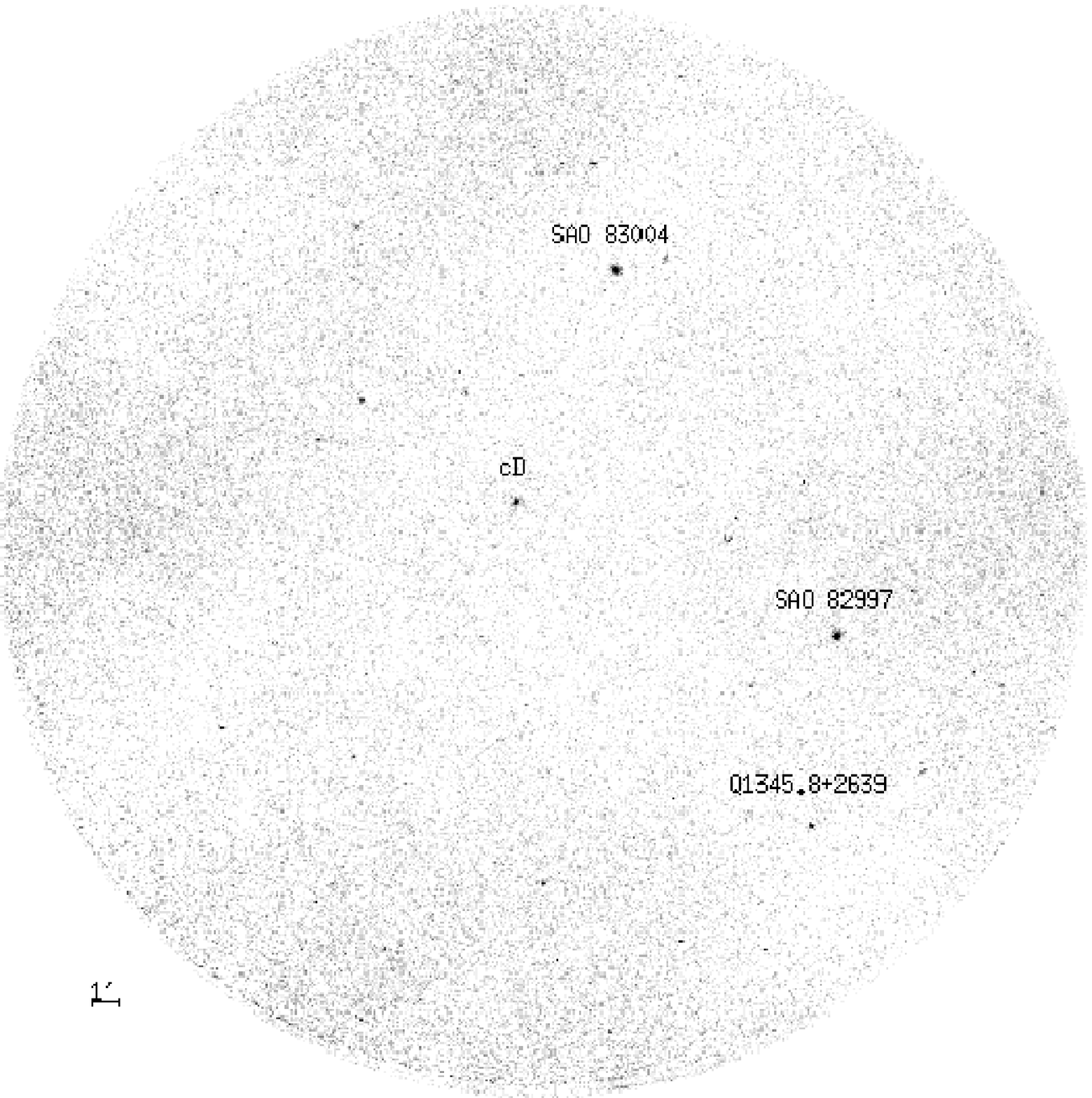]{ Full UIT image of the Abell 1795 cluster field centered on the
cD galaxy.  Other cataloged objects are labeled and discussed in
section~\ref{sec:others}.  The location and orientation of the HST WFPC
aperture used for its observation is overlaid for comparison.
\label{fig:cluster}}

\figcaption[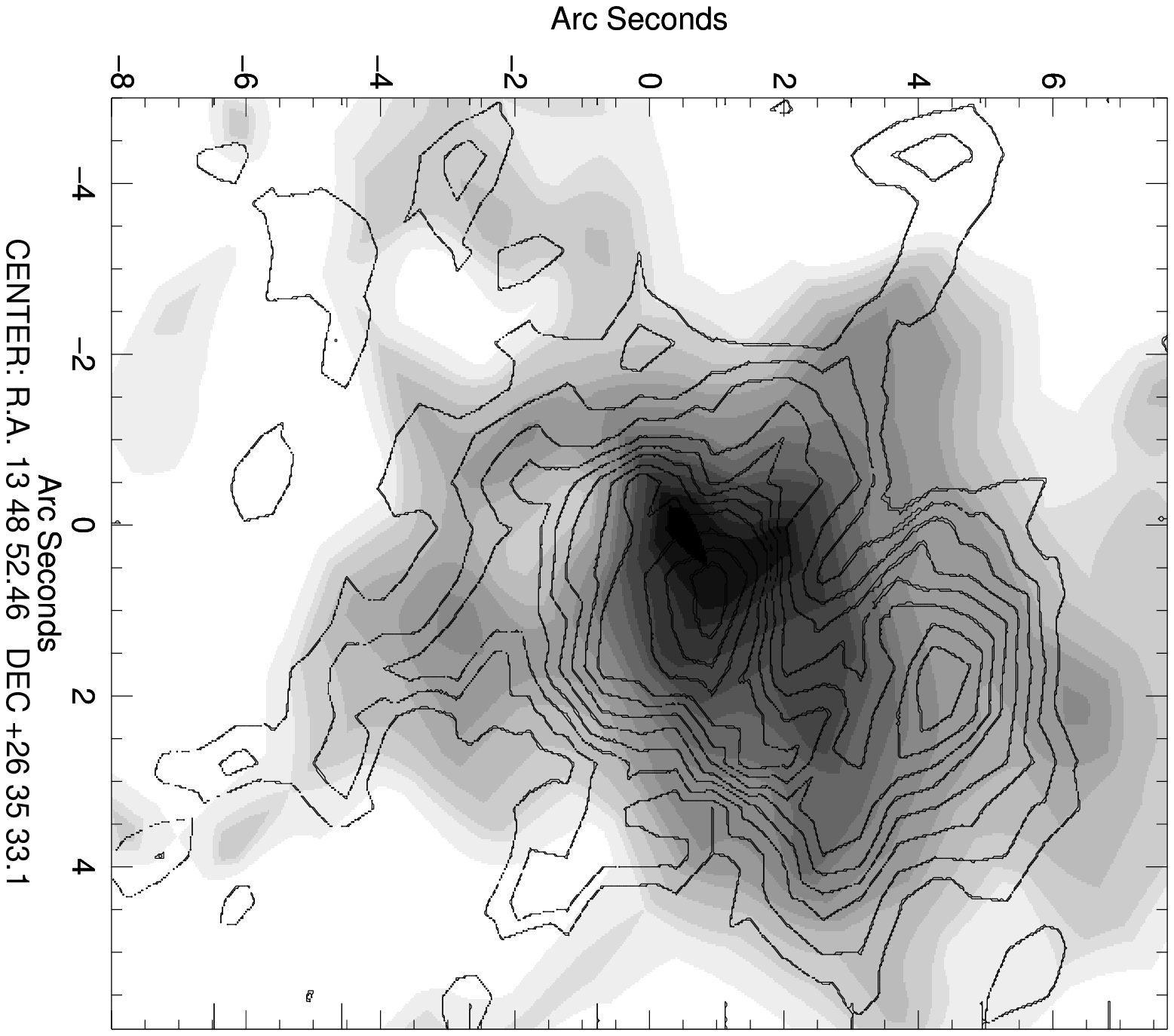]{H$\alpha$ contours from van Breugel, Heckman \& Miley (1984)
superimposed on a greyscale UIT image of the cD galaxy.  There is no far--UV
counterpart to the extended H$\alpha$ emission seen at $r=22$\arcsec
\label{fig:uitha}}

\figcaption[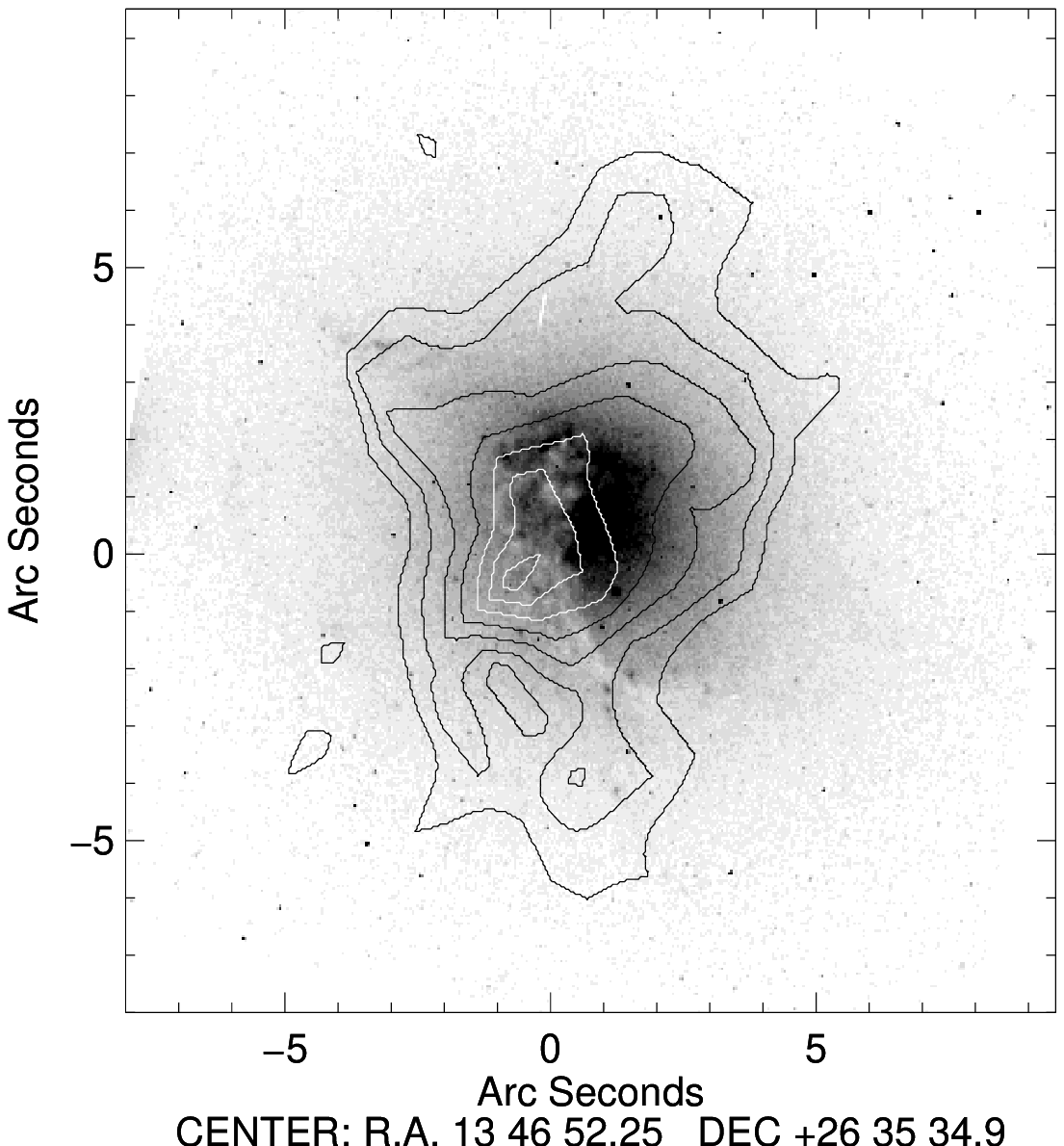]{ UIT far--UV (1520\AA\/ contours starting at $\sim3\sigma_{sky}$
increasing linearly by $1.1\times10^{-18}$ erg s$^{-1}$ cm$^{-2}$ \AA$^{-1}$
per contour superimposed on the HST F555W image.  The bulk of the cD
galaxy lies outside the PC chip boundaries.  Hence the
image is stretched to emphasize the innermost galaxy core. The 
contour colors are wrapped for clarity of contrast with the greyscale image.
\label{fig:hstuit}}

\figcaption[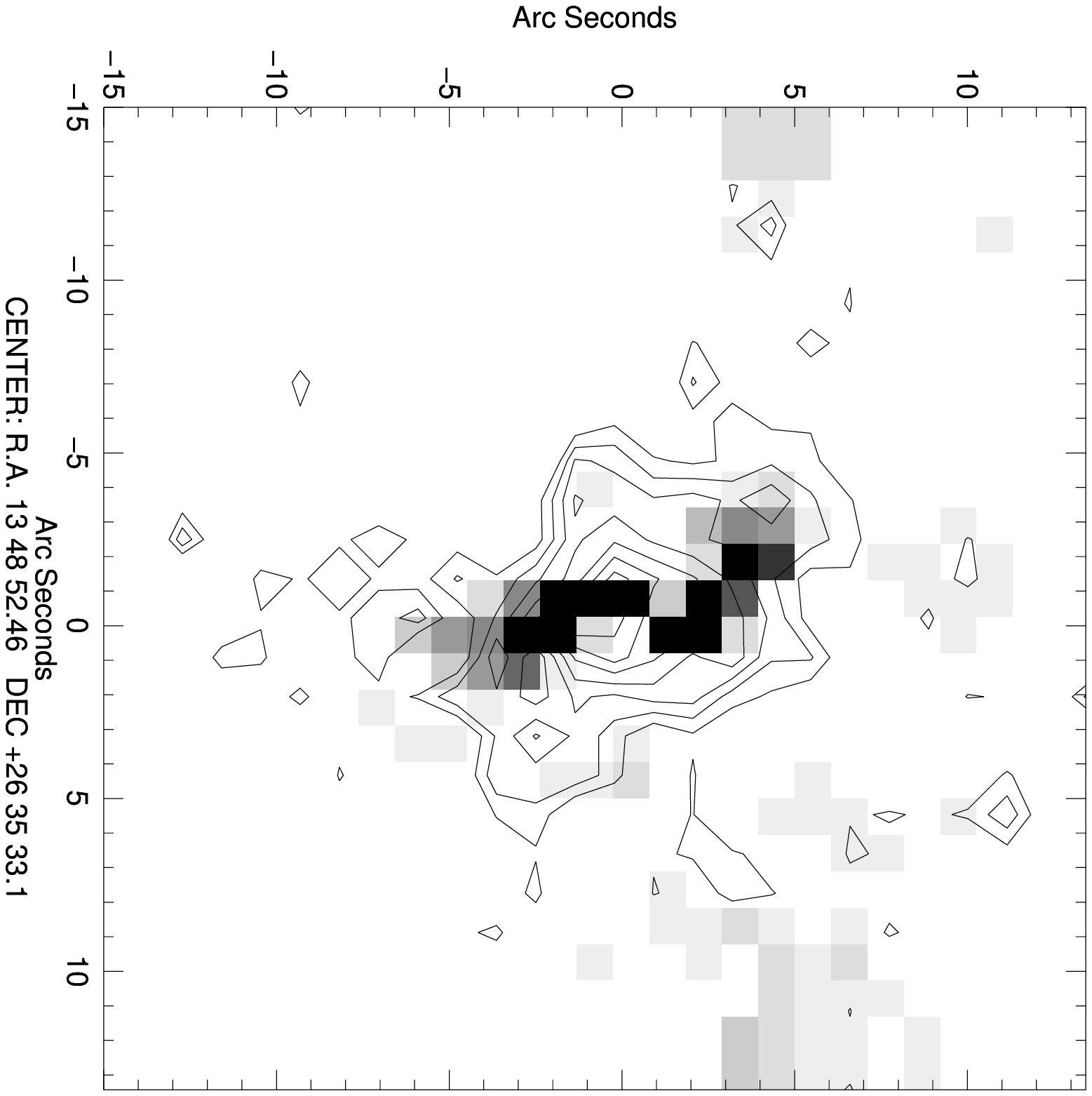]{UIT far--UV (1520\AA\/ contours starting at $\sim2\sigma_{sky}$
($7.2\times10^{-18}$ erg s$^{-1}$ cm$^{-2}$ \AA$^{-1}$ arcsec$^{-2}$ and incremented linearly
overlaid on the VLA 3.6 cm map from Ge \& Owen (1993).  The radio image has been
smoothed to the UIT resolution. \label{fig:uitvla}}

\figcaption[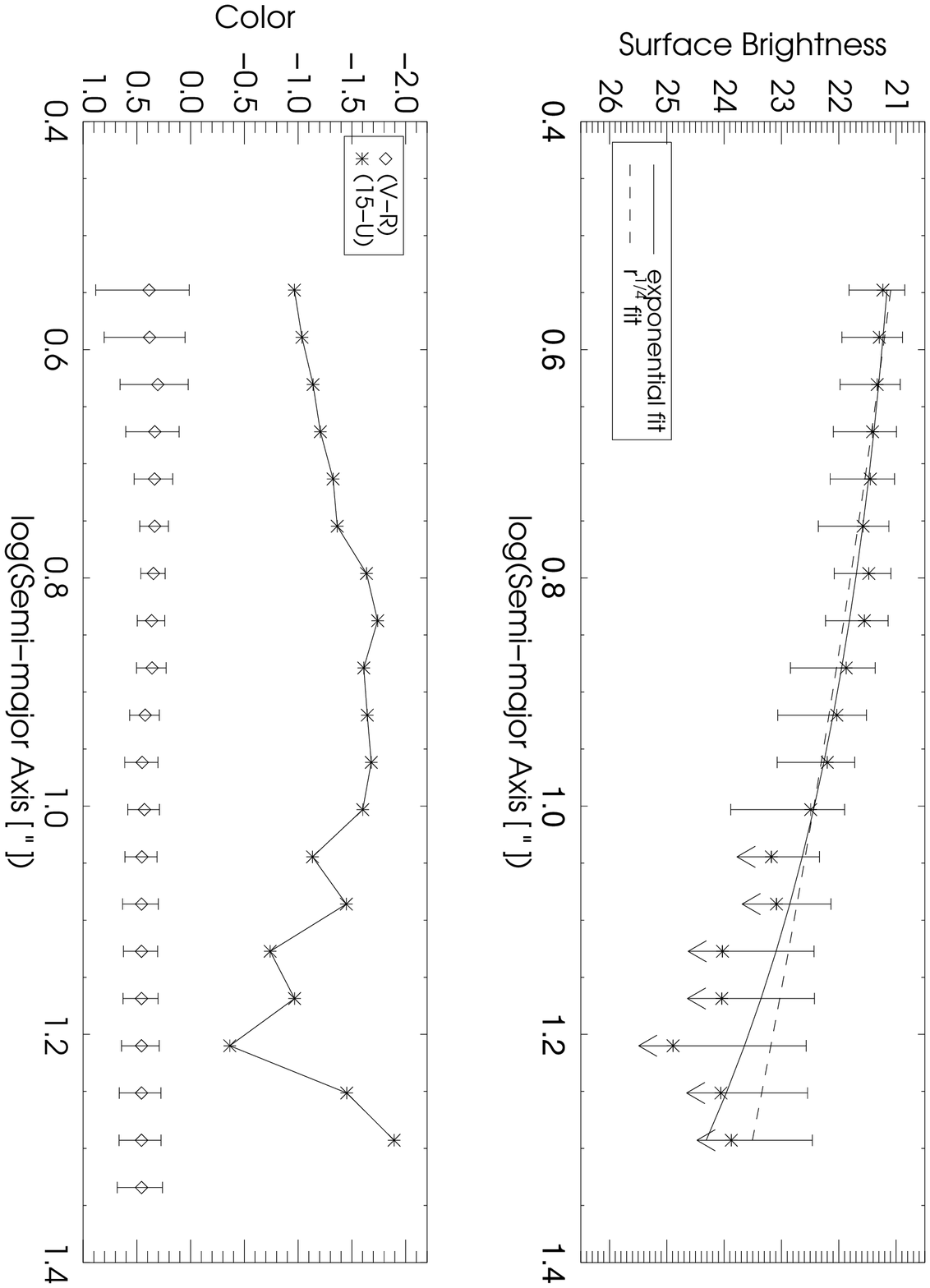]{Upper Panel: Surface brightness profile (magnitudes
arcsec$^{-2}$) in the 1520\AA\/ band along with best fitting disk and
elliptical galaxy models.  Lower panel: Color profiles for the galaxy over the
region defined by the surface brightness profile. \label{fig:profiles}}

\figcaption[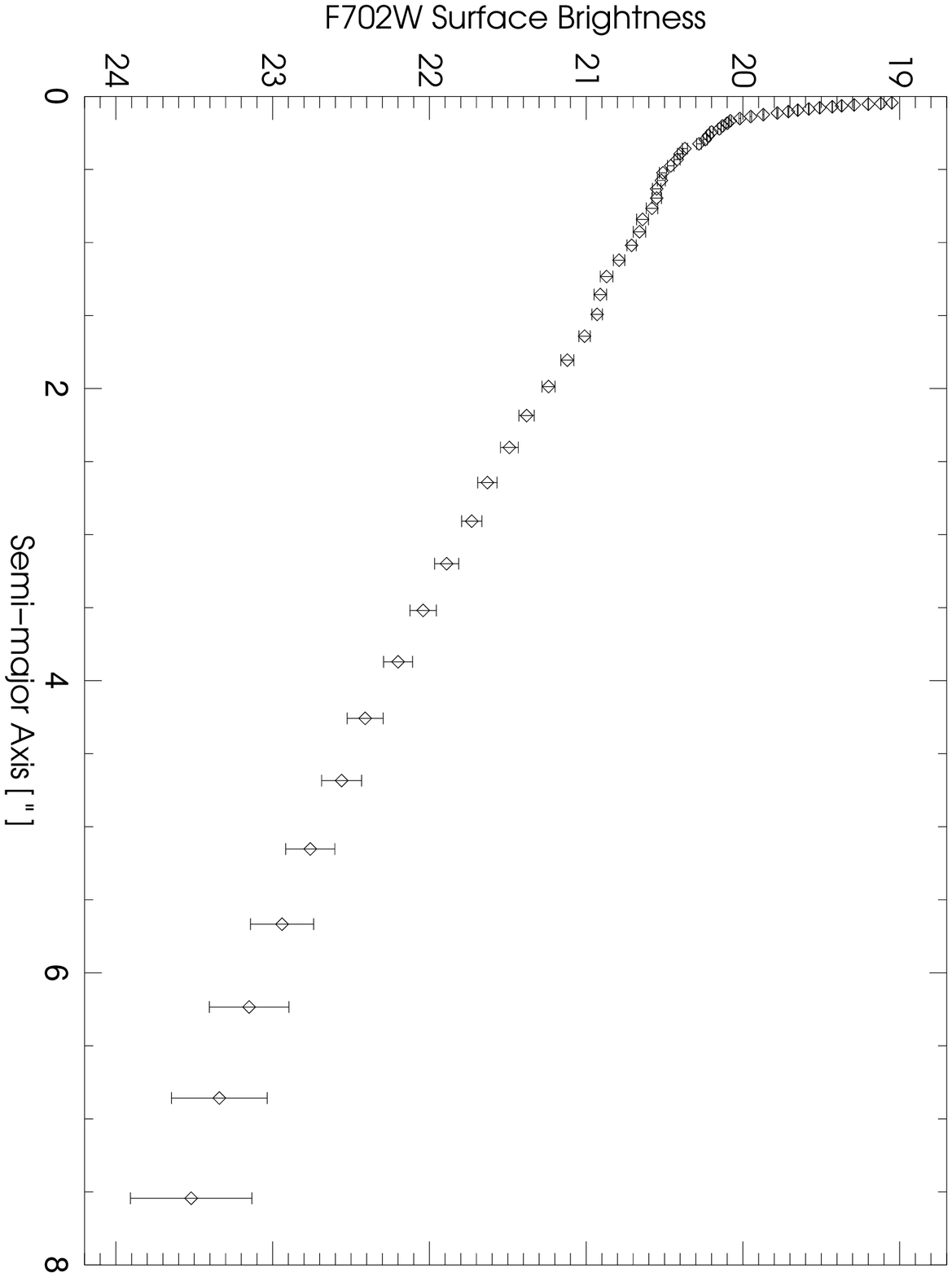]{Surface Brightness profile (mag arcsec$^{-2}$ for the HST 
F702W image of the central part of Abell 1795.  \label{fig:rprofile}}

\figcaption[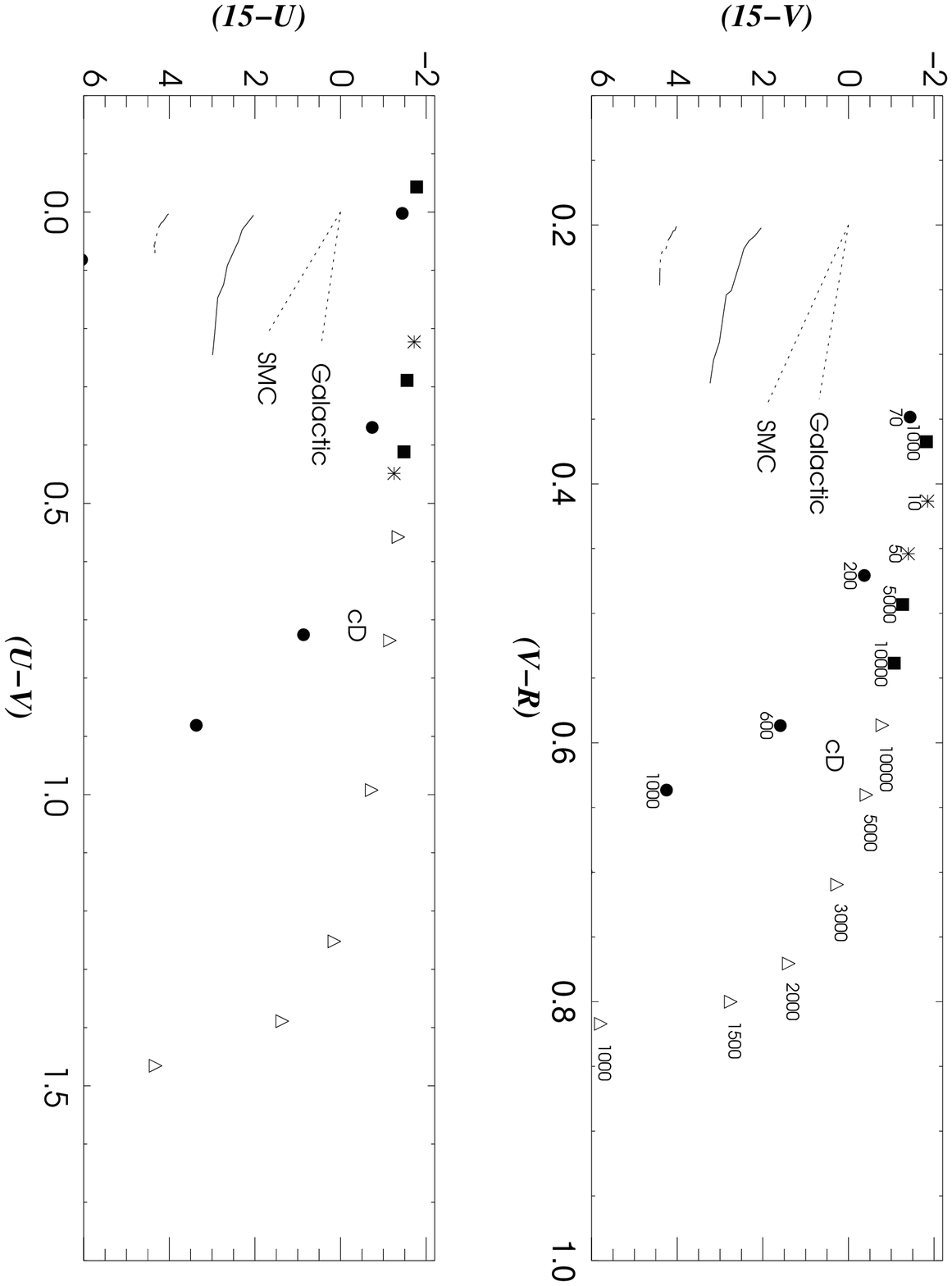]{Color--color diagrams showing the location of the cD along with
several models of star formation histories.  Symbols have the following
meaning: ($\bullet$) Single generation, instantaneous  bursts of star formation
with the age of the burst (in Myr) under the symbol, ($\triangle$) Single
generation, exponentially decaying star formation with the decay time (Myr) next to
the symbol, ($\Box$) Constant, continuous star formation, beginning at the time
listed under the symbol, ($\ast$) Single, 4 Myr old burst added to an old
population (see~\ref{sec:models}), (``cD'') Abell 1795 measured colors. 
Tracks indicating the effects of extinction are superimposed
(see~\ref{sec:dust}). The errors in the cD colors are smaller that the letters
used to label the point itself. \label{fig:c-cd}}

\clearpage
 
\plotone{figure1.eps}

\plotone{figure2.eps}

\plotone{figure3.eps}

\plotone{figure4.eps}

\clearpage

\plotone{figure5.eps}

\plotone{figure6.eps}

\plotone{figure7.eps}

\end{document}